\documentclass[onecolumn,showpacs,preprintnumbers,amsmath,amssymb]{revtex4}

\usepackage{epsf}
\usepackage{graphicx}  % Include figure files
\usepackage{dcolumn}   % Align table columns on decimal point
\usepackage{bm}        % bold math

\newcommand{\be}{\begin{equation}}
\newcommand{\en}{\end{equation}}
\newcommand{\bea}{\begin{eqnarray}}
\newcommand{\ena}{\end{eqnarray}}
\newcommand{\cp}{Chaplygin gas}
\begin{document}

\preprint{CAS-KITPC/ITP-100}

\title{A quantitative criteria for the coincidence problem}
 \author{Hongsheng Zhang}
% \affiliation{Department of Astronomy, Beijing Normal University, Beijing 100875, China} \affiliation{Korea Astronomy and Space Science Institute, Daejeon 305-348, Korea }
\affiliation{Department of Astronomy, Beijing Normal University, Beijing 100875, China}

 \author{Heng Yu}
\affiliation{Department of Astronomy, Beijing Normal University, Beijing 100875, China}

 \author{Zong-Hong Zhu}
 \email{zhuzh@bnu.edu.cn}
\affiliation{ Department of Astronomy, Beijing Normal University, Beijing 100875, China}
% \affiliation{Kavli Institute for Theoretical Physics China, CAS, Beijing 100190, China}

\author{Yungui Gong}
%\email{gongyg@cqupt.edu.cn} 
\affiliation{College of Mathematics and Physics, Chongqing University of Posts and Telecommunications, Chongqing 400065, China} 
\affiliation{Kavli Institute for Theoretical Physics China, CAS, Beijing 100190, China}

\date{ \today}

\begin{abstract}

 The cosmic coincidence problem is a serious challenge to dark energy model. We suggest a quantitative criteria for judging the severity of
 the coincidence problem. Applying this criteria to three different interacting models, including the interacting quintessence, interacting phantom, and
 interacting \cp~models, we find that the interacting \cp~model has a better chance to solve the coincidence problem.
 Quantitatively, we find that the coincidence index $C$ for the interacting~ \cp~model is
smaller than that for the interacting quintessence and phantom
models by six orders of magnitude.

\end{abstract}


\maketitle

\section{Introduction}

The type Ia supernovae observations in 1998 discovered that the
expansion of the universe is accelerating \cite{acce}. This
discovery of late time cosmic acceleration imposes a big challenge
and provides opportunities to gravitational and particle physics. To
explain the accelerated expansion, we can modify either the left hand 
side (modify the theory
of gravity) or the right hand side (add an exotic component with
negative pressure, dubbed as "dark energy") of Einstein equation.
Although lots of dark energy models have been proposed in the
literature, the nature of dark energy is still a mystery. For a
review of dark energy models, see for example \cite{review} and
references therein.

The simplest candidate of dark energy which is consistent with
current observations is the cosmological constant. Because of the
many orders of magnitude discrepancy between the theoretical
predication and the observation of vacuum energy, the origin of the
smallness of the value of the cosmological constant is unknown.
Furthermore, the cosmological constant faces the "coincidence
problem", namely, why the energy density of dark energy and dark
matter happens to be of the same order now? In terms of the
parameter $r$,
\be
 r\triangleq \frac{\rho_{dm}}{\rho_{de}},
 \en
where $\rho$ is the energy density, the subscript ${de}$ denotes
dark energy and ${dm}$ labels dark matter, the coincidence problem
says why $r$ becomes order of 1 now. If the current value of $r$
which is order of 1 is independent of initial conditions, then the
coincidence problem is alleviated. Therefore, the resolution of the
coincidence problem lies on the attractor solution. Along this line
of reasoning, the coincidence problem has been extensively studied
\cite{coin,pavon01,pavon03,pavon08}. For the quintessence model
without interaction, the attractor solution with acceleration is the
scaling solution with $r=0$, so the interaction between dark matter
and dark energy is proposed to get nonzero $r$ attractor solution.
Assuming that the interaction is turned on recently, the standard
radiation and matter dominated eras can be recovered and the
coincidence problem is alleviated because $r$ starts from order 1
unstable fixed point \cite{pavon03}.

To solve the coincidence problem, $r$ should not vary too much
through the whole history of the universe, in addition to having
attractor solution. Thus, during most of the history of the
universe, dark matter and dark energy evolve almost in the same way.
In other words, during the matter domination, dark energy behaves
like dark matter. The \cp~model satisfies this requirement
\cite{cp}. At early times, the \cp~model behaves like dark matter;
and it behaves like dark energy at late times. Furthermore, the
interacting \cp~model was shown to have the attractor solution with
nonzero $r$ in \cite{intergas}. So it can be used to solve the
coincidence problem. Here the \cp~model is used as a dark energy
model, not as a model unifying dark matter and dark energy.
Recently, a relationship between the Hubble parameter $H$ and the
ratio $r$ was found in \cite{prd2008}. Since the evolution of the
Hubble parameter $H(z)$ can be determined by observations
\cite{hzdata,hznew}, the evolution of the ratio $r$ can be derived.
In this letter, we discuss the evolution of $r$ for several
interacting dark energy models.

The letter is organized as follows. In section II, we study the
coincidence problem in the interacting quintessence (IQT) model, the
interacting phantom (IPT) model and the interacting Chaplygin gas
(ICG) model. We conclude the letter with some discussions in section
III.

\section{Three interacting models }
Due to the interaction between the two dark components, the
equations of motion for $\rho_{de}$ and $\rho_{dm}$ become
 \be
 \dot{\rho}_{de}+3H \gamma_{de} \rho_{de}=-\Gamma,
 \label{1st conti}
 \en
 and
 \be
 \dot{\rho}_{dm}+3H \gamma_{dm} \rho_{dm}=\Gamma,
 \label{2nd conti}
 \en
 where a dot stands for the derivative with respect to the cosmic time $t$, the barotropic index $\gamma$ is defined as
 \be
 \gamma=1+\frac{p}{\rho}=1+w,
 \en
 $w$ is the equation of state parameter, $\gamma_{dm}=1$,
and $\Gamma$ is the interaction term between dark energy and dark
matter. The origin of the interaction between dark energy and dark
matter is not clear and the interaction can be rather arbitrary.
Here we choose the interaction term \cite{pavon01,pavon03} \be
  \Gamma=3Hc(\rho_{dm}+\rho_{de}),
  \label{interterm}
  \en
where the constant $c$ is the coupling constant. A possible
mechanism for this interaction from scalar-tensor theory of gravity
is discussed in detail in \cite{intergas,inter,st}.

  \subsection{The evolution of $r$ in three interacting models}
  Combining Equations (\ref{1st conti}), (\ref{2nd conti}) and (\ref{interterm}),
  one obtains the variation of $r$ with respect to the cosmic time,
  \be
  \dot{r}=3Hr\left[\gamma-1+\frac{c
  (1+r)^2}{r}\right].
  \label{revolve}
  \en
The evolution of $r$ for the IQT and IPT models with constant
$\gamma$ has been discussed in \cite{pavon03,pavon08,prd2008}. The
attractor solution is \cite{pavon03,pavon08} \be
r_s=\frac{w+\sqrt{w^2+4c w}}{w-\sqrt{w^2+4c w}}, \en with the
condition $0<c<|w|/4$.
 For the ICG model,
$p_{ch}=-A/\rho_{ch}$, so
  \be
  \gamma_{ch}=1-\frac{H_0^4}{H^4}(1+r)^2A',
  \en
  where $A'=A(3H_0^2)^{-2}\kappa^4$ and $\kappa^2=8\pi G$. The  interacting \cp~model was studied in detail in \cite{intergas}.
  The attractor solution is $r_s=c/(1-c)$ with the condition
  $0<c<1$.

  With the help of Friedmann equation, the differential equation of $r$ with respect to $H$ reads \cite{prd2008},
  \be
  \label{drdh}
  \frac{dr}{dH}=-\frac{2r(1+r)}{H(\gamma+r)}\left[\gamma-1+\frac{\kappa^2\Gamma
  (1+r)^2}{9H^3r}\right].
  \en

\begin{figure}
 \centering
 \includegraphics[totalheight=2.95in, angle=0]{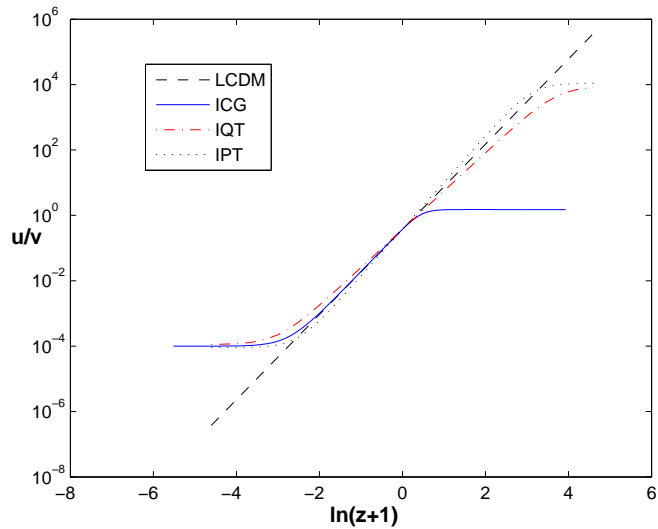}
 \caption{
 $r=u/v$ as a function of $\ln(1+z)$ in $\Lambda$CDM, IQT, IPT and ICG models. The initial conditions and the parameter are taken as:
 $u|_{z=0}+\Omega_b=0.3, v|_{z=0}=0.7$, and
 $\Omega_b=0.0389$. }
 \label{fig1}
 \end{figure}

 In terms of the dimensionless variable $u$, $v$ which are defined as
  \be
  u=(3H_0^2)^{-1}\kappa^2 \rho_{dm},
  \en
  \be
  v=(3H_0^2)^{-1}\kappa^2\rho_{de}, \en
 equations (\ref{1st conti}) and
  (\ref{2nd conti}) become
  \be
  \frac{du}{dx}=-3u+3c(u+v),
  \label{auto1}
  \en
  \be
  \frac{dv}{dx}=-3v(1+w_{de})-3c(u+v),
  \label{auto2}
  \en
  where $x=\ln a=-\ln(1+z)$. The Friedmann equation becomes
\be
  \frac{H^2}{H_0^2}=u+v+\Omega_b (1+z)^3,
  \en
where $\Omega_b$ is the fraction of the baryon energy density. The
above equations describe the evolution history of the universe. We
solve the above equations numerically for the $\Lambda$CDM, IQT,
IPT,  and ICG models, and the results are shown in Figure
\ref{fig1}. The parameters in the equations are taken as follows.
The Hubble constant $H_0=70$ kms${}^{-1}$Mpc${}^{-1}$ \cite{hznew}.
The baryon energy density $\Omega_b=0.0389$ \cite{wmap}. The
equation of state parameter $w_q=-0.9$ for the IQT model and
$w_p=-1.1$ for the IPT model. The coupling constant between dark
energy and dark matter $c_q=1.0\times 10^{-4}$ for the IQT model,
$c_p=1.0\times 10^{-4}$ for the IPT model, and $c_{Cp}=0.04$ for the
ICG model. Note that for the IQT model, observations require
$c<2.3\times 10^{-3}$ \cite{pavon08}, so we take take $c=1.0\times
10^{-4}$ \cite{prd2008}.

Note that there is no interaction between the baryon and dark energy
because the standard model of particle physics is well constrained
and the coupling between dark energy and ordinary matter is strongly
constrained by the solar system experiments \cite{will}. From figure
\ref{fig1}, it is clear that the ICG model solves the coincidence
problem because the ratio $r$ does not change much during the
evolution of the universe and the current ratio $r_s$ is the
attractor. In the  low redshift region, the three models almost
behave in the same way. But in
  the high redshift region, the ratio $r$ in the ICG model is effectively a constant,
  but it continually increases to a very
  high value for the IQT and IPT models.

 Currently there is no data available for the evolution of the ratio $r(z)$ yet.
 However, by solving equation (\ref{drdh}),
  we can express $H$ as a function of $r$, $H=H(r)$.
  For some special cases, especially for the quintessence and phantom models with constant
  equation of state parameter $w$,  the explicit function forms were
  obtained in \cite{prd2008}. Recently, the Hubble data $H(z)$ become available by
  using the age of the oldest galaxies \cite{hzdata,hznew}, and it has been
  used to constrain cosmological parameters \cite{coscon}.

  \begin{figure}
 \centering
 \includegraphics[totalheight=2.95in, angle=0]{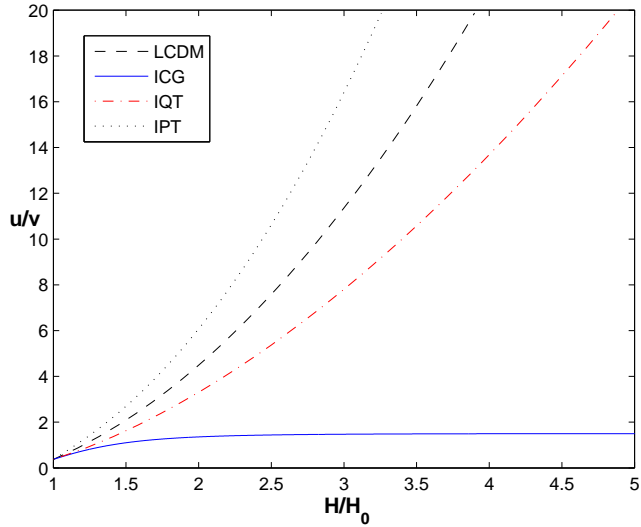}
 \caption{
 $r=u/v$ as a function of $\frac{H}{H_0}$  in $\Lambda$CDM, IQT, IPT and ICG models. The parameters are taken as the
 same as that of figure 1.}
 \label{fig2}
 \end{figure}

In Figure 2, we show the evolution of $H$ with respect to $r$. Form
Figure 2, one sees that in the ICG model, the ratio $r$ is almost a
constant for $H/H_0\geq 2$, while in the IQT and IPT models, $r$
increases with respect to $H$.

 \begin{figure}
 \centering
 \includegraphics[totalheight=2.95in, angle=0]{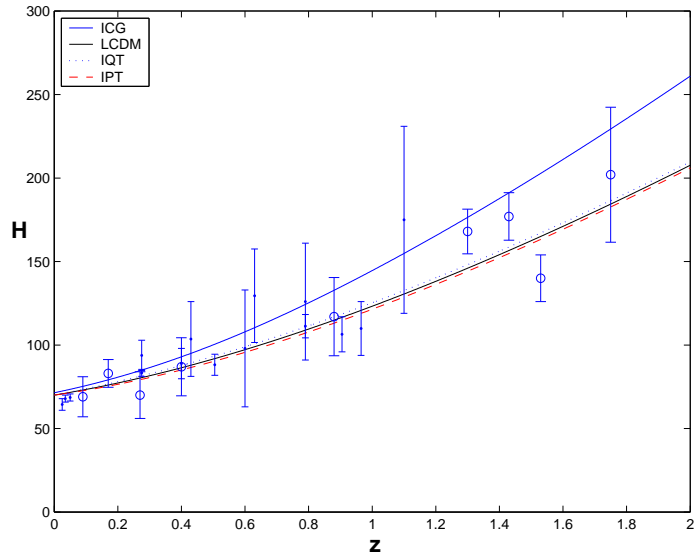}
 \caption{
 ${H}$ as a function of $z$ in $\Lambda$CDM, IQT, IPT and ICG modls. The parameters are taken as the
 same as that of figure 1. The 9 data of $H(z)$ (denoted by circles) with 1$\sigma$ error bars are taken from \cite{hzdata},
 while the other 15 data of $H(z)$ (denoted by points) with 1$\sigma$ error bars are taken from \cite{hznew}. }
 \label{fig2}
 \end{figure}

  In figure 3, we show the evolution of $H$ with respect to the redshift
  $z$.

  From Figures 1, 2, and 3, we see that the ICG model is a much better model
  for overcoming the coincidence problem. But in what degree? We
  need a quantitative criteria for the selection. We shall discuss
  this point in the next subsection.

   \subsection{A quantitative criteria for coincidence}

   The interacting models are often invoked to solve the
   coincidence problem. As we have mentioned before, the attractor
   behavior in the interacting model is a necessary ingredient to overcome the coincidence problem.
 For the attractor solution, the ratio $r$ is independent of the initial conditions, and therefore greatly
 softens the coincidence problem. For such models, the
  ratio only depends on the parameters of the model,  but is independent
  of the initial conditions. Therefore there is no problem of fine-tuning the initial conditions. However,
  we still need to choose appropriate model parameters to get the right value of $r_s$. In this sense, there
  exists the problem of fine-tuning the parameters. If the ratio $r$ does not change much during the whole history
  of the universe for most of the parameters, then the coincidence problem can be solved.
  Based on these considerations, we suggest a quantitative criteria for judging the severity of
  the coincidence problem.  We divide the coincidence problem into two smaller problems. The first is
  the coincidence between the value of $r_e$ at early time and that of $r_0$ at present. The second
  is the coincidence between $r_0$ and the attractor value (if it exists) $r_f$. To describe the coincidence problem quantitatively,  we
define the
  indices of early coincidence and late coincidence, respectively,
  \be
  C_e\triangleq \frac{r_e}{r_0},
  \en
  \be
  C_f\triangleq \frac{r_f}{r_0}.
  \en
  The closer to 1 $C_e$ or $C_f$ is,
   the better the coincidence problem is overcome. However we can not define the index of
   the coincidence as $C_e C_f$, since $C_e$ and $C_f$ maybe
   varies in the opposite direction, which
   makes the coincidence problem less severe. For
   example, if $C_e=10^{10}$ and $C_f=10^{-10}$, the $C_e C_f =1$. Therefore, we introduce a function,
   \be
   F(x)=\left\{
       \begin{array}{l l}
       x,& x>1 \\
       1/x,& 0\leq x\leq 1
       \end{array}
       \right.
   \en
By using the function $F(x)$, we can
  define a proper index of coincidence $C$ in the whole history of the
  universe as follows,
  \be
  C\triangleq F(C_e)F(C_f).
  \en
     which evades the above mentioned problem.

    We must take a standard epoch
     in the early universe to determine $C_e$. A most natural choice is $z\to \infty$.
     However, we should not expect
     a phenomenological model can describe the whole evolution
     history of the universe, especially the physics of the very early universe. In this letter, we take
     $z=100$ as ``early universe".

  From equation (\ref{revolve}), the fixed point of $r$ is
  \be
  w+c\frac{(r_s+1)^2}{r_s}=0,
  \en
  which yields,
  \be
  r_s=-(1+\frac{w}{2c})\pm \sqrt{\frac{w}{c}+\frac{w^2}{4c^2}}.
  \en
  Since $r$ should be real, so
  \be
  \frac{w}{2c}\geq 0,
  \label{nega}
  \en
  or
  \be
  \frac{w}{2c}\leq -2.
  \label{posi}
  \en
  The condition (\ref{nega}) leads to negative $r_s$,
  which is not physical. Hence, for the existence of the fixed point, the parameters $w$ and $c$ are required to satisfy
  the condition (\ref{posi}). If $w$ is a function of cosmic time, like the case in the \cp~ model, then in
  (\ref{posi}) we take the the value of $w$ as $\lim_{z\to -1} w$.
  For the ICG model, we take $c=0.06,~ A'=0.4$, which is
 consistent with observations \cite{intergas}. For the IQT model, we take $c=1.0\times 10^{-4}$ \cite{prd2008}.

 \begin{table}
 \begin{center}
 \begin{tabular}{ccccc}
 \hline\hline
 $~~~~~~~$ &  ~~~~~~IQT &~ IPT~~~&~~~~~ICG\\ \hline
  $C_e$ & ~~~~~$10^4$ & $10^4$ & 0.149\\ \hline
 $C_f$ &~~~~~ $10^{-4}$ & $10^{-4}$ & 0.149\\
 \hline
 $C$ & ~~~~~$10^{8}$ & $10^8$& $45$\\ \hline
 \hline
 \end{tabular}
 \end{center}
 \caption{\label{allow}
 Coincidence indices for the IQT, IPT and ICG models.}
 \end{table}

We apply this criteria to the three interacting models. The result
is shown in table I.
 From table I, it is clear that the ICG model evades
 the coincidence problem. Quantitatively, the coincidence index $C$ for the interacting~ \cp~ model is
smaller than that for the interacting quintessence and phantom
models by six orders of magnitude.

  %%%%%%%%%%%%%%%%%%%%%%%%%%%%%%%%%%%%%%%%%%%%%%%%%%%%%%%%%%%%%%%

\section{Conclusion and discussion}

In order to study the coincidence problem, we propose a quantitative
criteria for the determination of the severity of the
 coincidence problem.

 First, we study the evolution of the ratio $r$ of dark matter to dark energy in the
IQT, IPT, and ICG models. Though presently we
 do not have the information about $r(z)$, we can express $r$ with the
 Hubble parameter $H$. We found the evolutions of $r$ with respect to the redshift $z$ and the Hubble parameter $H$
 for the three interacting models.
 We also found the evolution of the Hubble parameter $H(z)$ and compared it with the observational data.
 We find that the ICG model solves the coincidence problem.

 Through a detailed analysis of the coincidence problem, we
  suggest a quantitative criteria for the determination of the severity of the coincidence problem.
  Applying this criteria to the IQT, IPT, and ICG models, we find that the ICG model solves the coincidence problem
  and the coincidence index $C$ for the interacting~ \cp~ model is
smaller than that for the interacting quintessence and phantom
models by six orders of magnitude..

 {\bf Acknowledgments.}
 Z-H Zhu  was supported
by the National Science Foundation of China under
 the Distinguished Young Scholar Grant 10825313,
 the Key Project Grant 10533010,
and by the Ministry of Science and Technology national
 basic science Program (Project 973) under grant No. 2007CB815401.
 YG Gong was supported by NNSFC under Grant No. 10605042.

\end{document}